\documentclass[twocolumn,prl,aps,showpacs]{revtex4} \voffset
0.4in

\usepackage{graphics}
\usepackage{dcolumn}
\usepackage{bm}
\usepackage{epsfig}
\usepackage{pictex}

\begin{document}

\title{Non-local Coulomb correlations in metals close to a charge order insulator transition}
\author{Jaime Merino}
\affiliation{Departamento de F\'isica Te\'orica de la Materia
Condensada, Universidad Aut\'onoma de Madrid, Madrid 28049, Spain
\\}
\date{\today}
\begin{abstract}
The charge ordering transition induced by the nearest-neighbor
Coulomb repulsion, $V$, in the $1/4$-filled extended Hubbard model
is investigated using Cellular Dynamical Mean-Field Theory. We
find a transition to a strongly renormalized charge ordered Fermi
liquid at $V_{\text{CO}}$ and a metal-to-insulator transition at 
$V_{\text{MI}}>V_{\text{CO}}$. Short range antiferromagnetism occurs
concomitanly with the CO transition. Approaching the charge ordered insulator, $V\lesssim
V_{\text{MI}}$, the Fermi surface deforms and the scattering rate of electrons develops 
momentum dependence on the Fermi surface.  
\end{abstract}
\maketitle

Strongly correlated electronic systems in low dimensions exhibit
novel quantum states of matter with unconventional electronic
properties. Among the fascinating phenomena observed in these
systems is the Mott metal-insulator transition induced by the local
Coulomb repulsion energy\cite{Mott} which occurs, for instance, in
transition metal oxides and layered organic superconductors
\cite{Ishiguro,Kanoda}. Important progress in the description of
the Mott transition\cite{Imada1} has been achieved with the
development of dynamical mean-field theory (DMFT)\cite{Georges}, a
local theory which predicts that close to the Mott insulator (MI), a 
quasi-particle peak and Hubbard bands coexist in agreement with photoemission
spectra of transition metal oxides\cite{Kotliar} and optical
spectra of $\kappa$-(BEDT-TTF)$_2$X \cite{Dressel}. 
Non-local  correlations can modify this scenario
opening a pseudogap in the spectral density as Cellular Dynamical
Mean-Field Theory (CDMFT) \cite{Kyung} and the Dynamical Cluster 
Approximation (DCA) \cite{Maier} have found consistent 
with the pseudogap phase of high-T$_c$ superconductors.

The $1/4$-filled family of layered organic molecular superconductors of the $\alpha$, $\beta''$ 
and $\theta$-(BEDT-TTF)$_2$ X types display rich phase diagrams\cite{Seo}
and anomalous metallic behavior\cite{Natalia} such as the
Drude peak absence in the optical conductivity of $\beta''$-(BEDT-TTF)$_2$
SF$_5$CH$_2$CF$_2$SO$_3$\cite{Dong}.
In these quasi-two-dimensional systems  
both the on-site Coulomb repulsion, $U$, and the
nearest-neighbor Coulomb repulsion $V$ are substantial.   
In spite of its importance, the electronic properties 
of metals close to a Coulomb-driven charge ordered insulator (COI) transition 
remain poorly understood. Taking the
extended Hubbard model as the minimal strongly correlated model which 
includes the competition between itinerancy and localization due to charge ordering       
we study the properties of metals close to a COI.  
An important theoretical challenge is to  
understand the evolution of the spin degrees of freedom across the 
charge ordering transition and, in particular, 
whether spin and charge order together or not.

In this Letter, we describe how the COI transition driven by the
Coulomb repulsion, $V$, in the extended Hubbard model occurs based
on the CDMFT approach, which is a strong coupling method which includes both
on-site and short range non-local correlations. We
find a transition to a strongly renormalized charge ordered metal 
(COM) at $V_{\text{CO}}$ and a metal-insulator transition at $V_{\text{MI}}$. 
We find that the spectral density is suppressed on some 
regions of the Fermi surface due to the momentum dependence 
developed in the electron scattering rate close to the COI.   
Concomitanly with the CO transition an antiferromagnetic (AF) interaction 
between spins at charge rich sites of the COM is dynamically 
generated by the Coulomb interaction, $V$.

The simplest strongly correlated model that captures the
competition between charge ordering and the homogeneous metal is an
extended Hubbard model on a square lattice at $1/4$-filling:
\begin{eqnarray}
H = &-t& \sum_{\langle ij \rangle,\sigma} (c^\dagger_{i \sigma}
c_{j \sigma} + c^\dagger_{j \sigma} c_{i \sigma}) + U \sum_{i}
n_{i\uparrow} n_{i\downarrow}
\nonumber \\
&-&\mu \sum_{i \sigma} c^{\dagger}_{i\sigma} c_{i \sigma}
+\sum_{\langle ij \rangle} V_{ij} n_i n_j, \label{ham}
\end{eqnarray}
where $\sigma$ is the spin index and $c^\dagger_{i \sigma}$ and
$c_{i \sigma}$ denote the electron operators. $t$ is the
transfer integrals between nearest-neighbor sites of the square lattice. 
The Coulomb interaction is such that: $V_{ij}=V$ if $i$ and $j$ are
nearest neighbor sites and $V_{ij}=0$ otherwise. The system
with no interactions ($U=V=0$) has a dispersion with bandwidth
$W=8t$. We fix $U=10t$ and analyze the CO transition due to $V$ at
1/4-filling.

\begin{figure}
\begin{center}
\epsfig{file=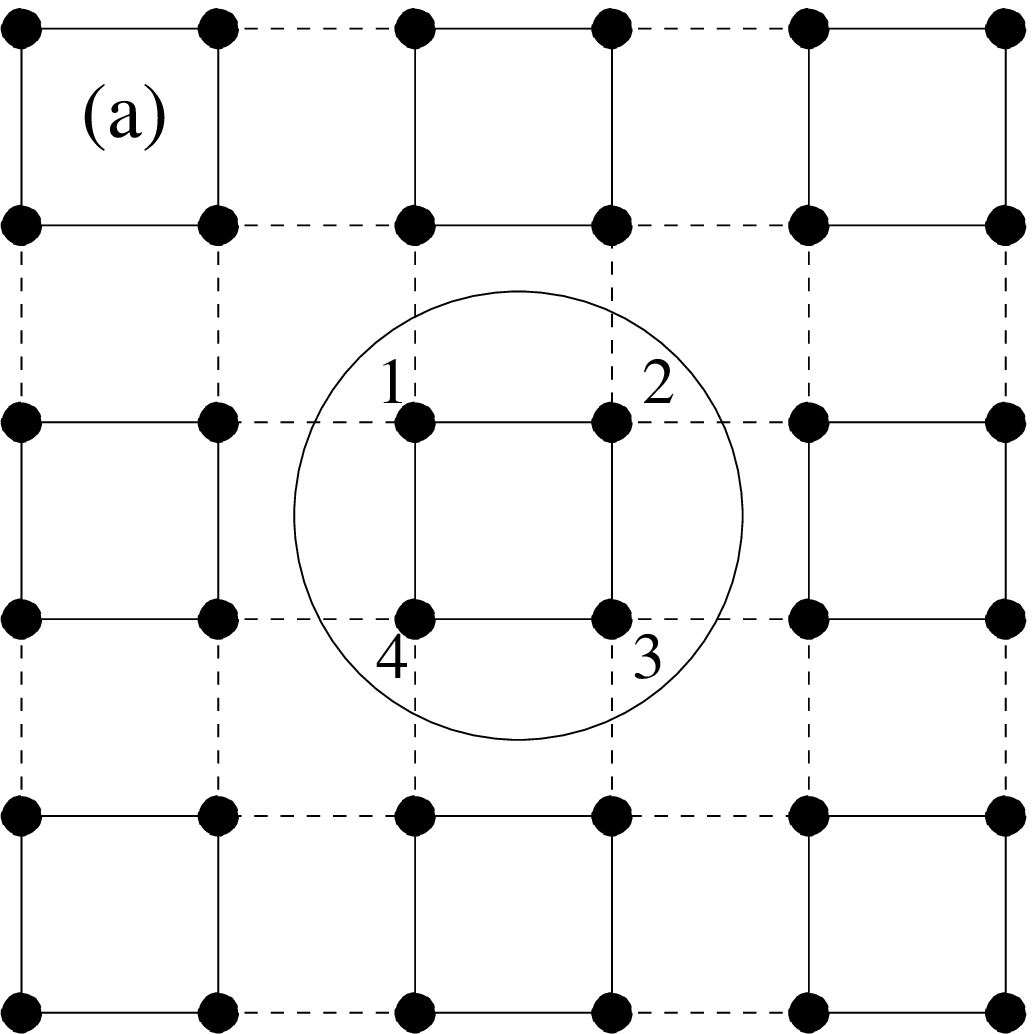,width=3.5cm,angle=0,clip=}
\epsfig{file=fig1b.eps,width=3.5cm,angle=0,clip=} \vskip 0.5cm
\epsfig{file=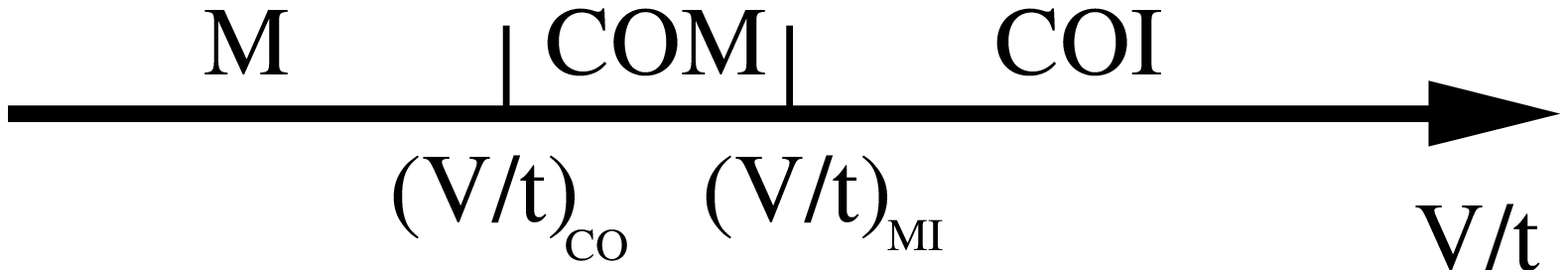,width=7.0cm,angle=0,clip=}
\end{center}
\caption{(Color online) Ground state phase diagram of the 1/4-filled extended
Hubbard model on the square lattice from CDMFT calculations. In
(a) the tiling of the two-dimensional lattice used in CDMFT is
shown whereas in (b) the $V$-dependence of site occupations,
$n_i$, is plotted for $U=10t$. } \label{fig1}
\end{figure}

In the CDMFT method, the infinite lattice is divided into
identical clusters of size $N_c=4$ which are treated exactly as
shown in Fig. \ref{fig1} (a). The Coulomb interaction 
between electrons in different clusters is treated at the mean-field level.
The quantum cluster is embedded in a
non-interacting bath which is solved self-consistently. Results are
presented for paramagnetic phases unless explictly stated. The
effective Anderson model reads
\begin{eqnarray}
H &=&\sum_{\alpha k} \epsilon^{\alpha}_k n_k +\sum_{i\sigma} \epsilon_{i} n_{i
\sigma} +\sum_{\alpha ki\sigma} \Gamma^{\alpha}_{ki} (c^+_{k \sigma} c_{i \sigma}
+ c. c.)
\nonumber \\
&+&U\sum_i n_{i \uparrow} n_{i\downarrow} +\sum_{ij} t_{ij}(c^+_{i
\sigma} c_{j \sigma} +  c^+_{j \sigma} c_{i \sigma}) + \sum_{ij}
V_{ij} n_i n_j,
\nonumber \\
\end{eqnarray}
where $t_{ij}=-t$ denote hopping matrix elements
between neighbor sites inside the cluster with $i$ and $j$ running from 1 to
Nc. Following Ref. [\onlinecite{Kancharla}], bath levels are grouped
into $\alpha$ subsets. Correspondingly the energies,
$\epsilon^{\alpha}_k$, refer to the $k$ energy level of the
$\alpha$ subset.  The bath-cluster couplings between a bath level
$k$ of bath $\alpha$ and the cluster site $i$ are:
$\Gamma^\alpha_{k i}$. In practice, we truncate the bath to
$N_b=8$ sites which are grouped into two subsets: $\alpha=1,2$
with $k=1,..,4$ states each. We take $N_c=4$ cluster sites. Lattice
symmetries (allowing for checkerboard charge ordering) reduce the bath
parameters to 4 independent bath levels: $\epsilon^{\alpha}_k =
\epsilon^{\alpha}_{k+2}$, and hybridization bath parameters:
$\Gamma^{\alpha}_{ki} = \delta_{ki} \Gamma^{\alpha}_i$, 
with $\Gamma^{\alpha}_i=\Gamma^{\alpha}_{i+2}$. 
Thus, we use 4 independent bath levels and 4 independent hybridization 
parameters to fully describe the bath. 
After extracting the self-energy of
the cluster, $\Sigma_{ij}(i \omega_n)$, for an arbitrary choice of
$\epsilon_{i}$'s and $\Gamma_{k i}$'s the lattice Greens function
is found by imposing the superlattice periodicity:
\begin{equation}
G_{ij}(i \omega_n)=\sum_{\bf K} ( (i \omega_n +\mu-\sum_k V_{ik}
\langle n_k \rangle )\delta_{ij}-t_{ij}({\bf K})-\Sigma_{ij}(i \omega_n) )^{-1},
\label{greens}
\end{equation}
where $t_{ij}({\bf K})$ is the Fourier transform of the cluster
hopping amplitudes and ${\bf K}$ the momentum wavevector in the
reduced Brillouin zone of the superlattice. Matsubara frequencies,
$\omega_n=(2n+1)\pi/\beta$, with an inverse temperature: $\beta=200/t$ are used. 
$\langle n_k \rangle$ is the
average electronic occupation in nearest-neighbor sites in neighboring clusters.  
The one-dimensional version of model (\ref{ham}) has been previously
studied with CDMFT\cite{Bolech}. The two-dimensional Hubbard model
on a frustrated square lattice close to half-filling (with small
degree of frustration), leads to an intricate competition between
AF, Mott insulator, unconventional
superconductivity and metallicity\cite{Lichtenstein}. 


The resulting phase diagram of the extended Hubbard model based on 
CDMFT is shown in Fig. \ref{fig1}. As $V$ is increased the
system undergoes a sharp transition from an homogeneous 
metal (M) to a strongly renormalized charge ordered metal (COM) at
$V_{\text{CO}} \approx 1.3t$. The jump in the order parameter of the CO transition,
$n_{12}=\langle n_1 \rangle - \langle n_2 \rangle $ is evident   
from Fig. \ref{fig1} (b). Initially, for $V < V_{\text{CO}}$, the system is
weakly correlated as the quasiparticle weight $Z_i=(1-\partial
\text{Im}\Sigma_{ii}(i \omega_n)/\partial \omega_n|_{\omega=0} )$ in
charge rich ($Z_1$) and charge poor ($Z_2$) cluster sites is $Z_1
\sim Z_2=0.7$ consistent with slave-boson theory\cite{McKenzie}. The weakly correlated metal
becomes a strongly renormalized Fermi liquid once  
charge ordering sets in with $Z_1=0.1$ at $V=1.5t$ indicating 
localization of electrons at every other site of the 
lattice due to CO. At $V_{\text{MI}}\approx 2t$
a gap opens in the spectral DOS indicating a MI transtion.
Previous direct Lanczos calculations on 20-site clusters predict a MI transition 
at $V_{\text{MI}} \approx 2.2t$ and indicated a COM for $V>V_{\text{CO}} \approx 1.5t$
in remarkable agreement with our present results despite the small size
of the cluster used within CDMFT.

\begin{figure}
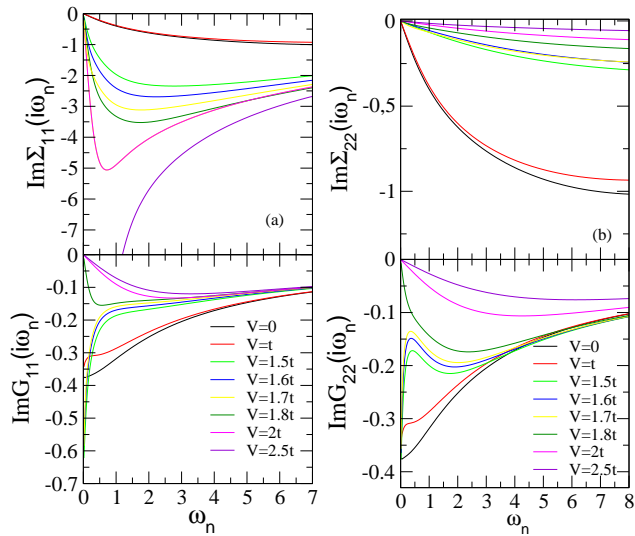

\begin{center}
\epsfig{file=fig2a.eps,width=4.1cm,angle=0,clip=}
\epsfig{file=fig2b.eps,width=4.1cm,angle=0,clip=}
\end{center}
\caption{(Color online) Enhancement of on-site self-energies across the CO and MI
transitions with $V$. In (a) $\Sigma_{11}(i \omega_n)$ and
$G_{11}(i\omega_n)$ which correspond to the charge-rich sites are
plotted and in (b) the same for the charge-poor
sites.} \label{fig2}
\end{figure}

\begin{figure}
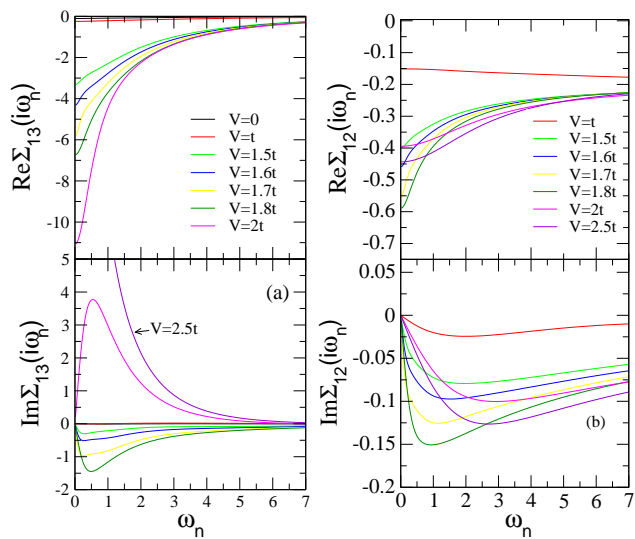

\begin{center}
\epsfig{file=fig3a.eps,width=4.0cm,angle=0,clip=}
\epsfig{file=fig3b.eps,width=4.2cm,angle=0,clip=}
\end{center}
\caption{(Color online) Dynamically generated off-site interaction in the COM.
Non-diagonal self-energies are plotted with $V$. Self-energies
$\Sigma_{ij}(i \omega_n)$ between two next-nearest neighbor charge
rich sites are plotted in (a) while between nearest-neighbor 
sites are displayed in (b).} \label{fig3}
\end{figure}

Dynamical and short range correlation effects  
are analyzed based on cluster quantities: $G_{ij}(i \omega_n)$
and $\Sigma_{ij}(i\omega_n)$. Figs. \ref{fig2} and \ref{fig3} show how diagonal 
and off-diagonal self-energies are 
enhanced for  $V>V_{\text{CO}}$ being most significant between 
charge rich sites (see $\Sigma_{13}(i\omega_n)$ in Fig. \ref{fig3}).
Consistent with the self-energy enhancement, Green's functions are 
suppressed at finite $\omega_n$ for $V>V_{\text{CO}}$. For $V \gtrsim 2t$,  
Im$G_{ii}(i\omega_n \rightarrow 0) \propto -\omega_n$, displaying 
insulating behavior \cite{Ansgar}. This contrasts to the metallic-like behavior
of Im$G_{ii}(i\omega_n)$ for $V<V_{\text{MI}}$ indicating a 
MI transition at $V_{\text{MI}}=2t$. Fig. \ref{fig3} shows  
Im$\Sigma_{ij}(i \omega_n \rightarrow 0) \rightarrow 0$ indicating that 
the opening of the gap is due to the large 
Re$\Sigma_{13}(i 0^+)$ developing close to the COI (see Fig. \ref{fig3}).
This interaction induced by Coulomb interaction produces a band splitting at 
the Fermi energy\cite{Held}. Note that as LRO-AF is not allowed, the 
opening of the gap at $V_{\text{MI}}$ can only be due to short range correlations.
We also find that at $V \approx 2.5t$, Im$\Sigma_{11}(i0^+)$ and
Im$\Sigma_{13}(i0^+) $ diverge.  

\begin{figure}
\begin{center}
\epsfig{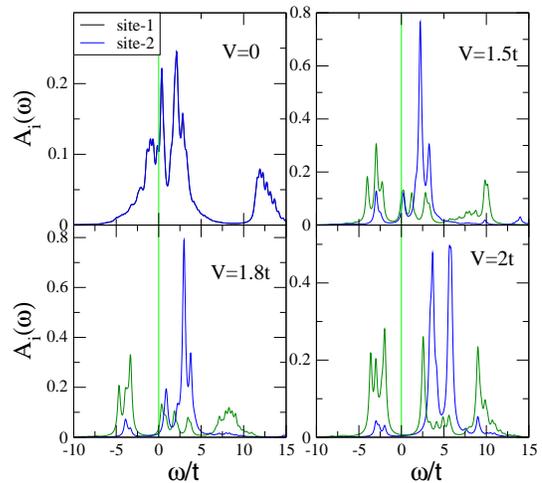}
\end{center}
\caption{(Color online) Evolution of the DOS with the Coulomb repulsion, $V$, 
across the CO transition.  $A_{i}(\omega)$ is plotted for different $V$ 
showing a gap opening at the COI transition $V_{\text{MI}}=2t$. 
The vertical line denotes the position of the chemical potential.
} \label{fig4}
\end{figure}

\begin{figure}
\begin{center}
\epsfig{file=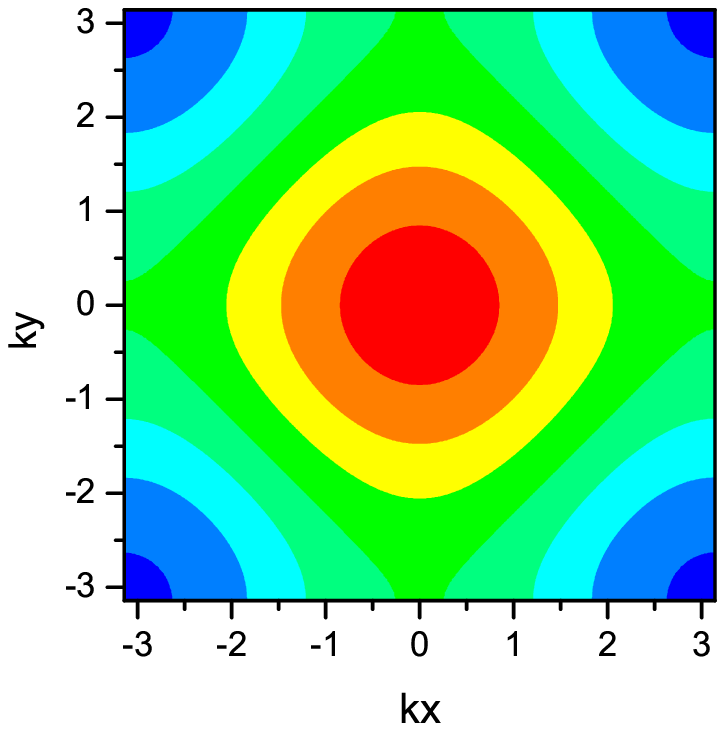, width=4.cm,angle=0,clip=}
\epsfig{file=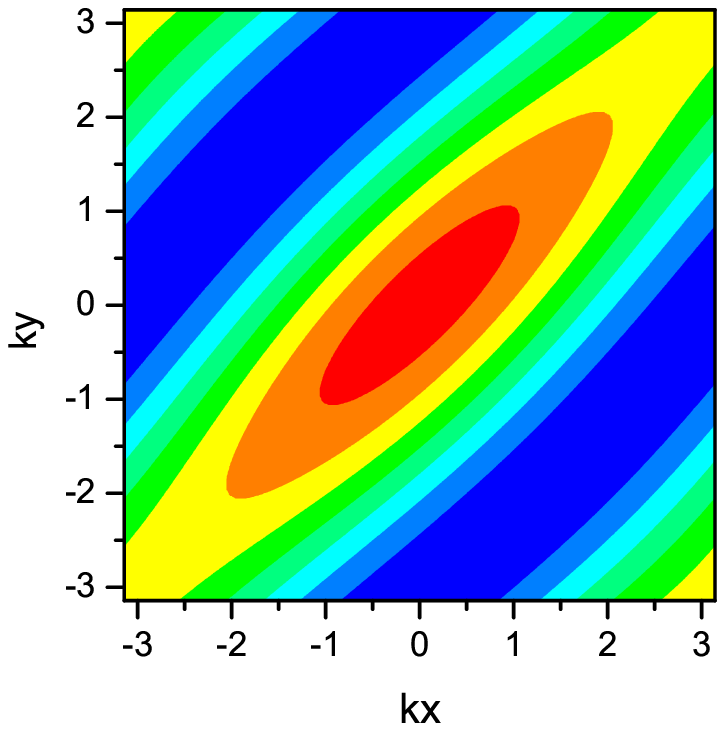, width=4.cm,angle=0,clip=}
\epsfig{file=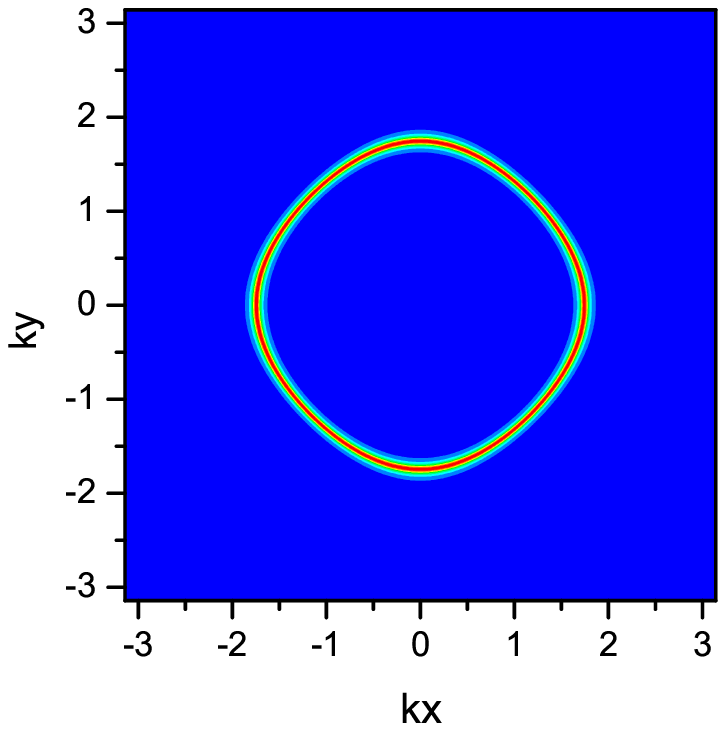, width=4.cm,angle=0,clip=}
\epsfig{file=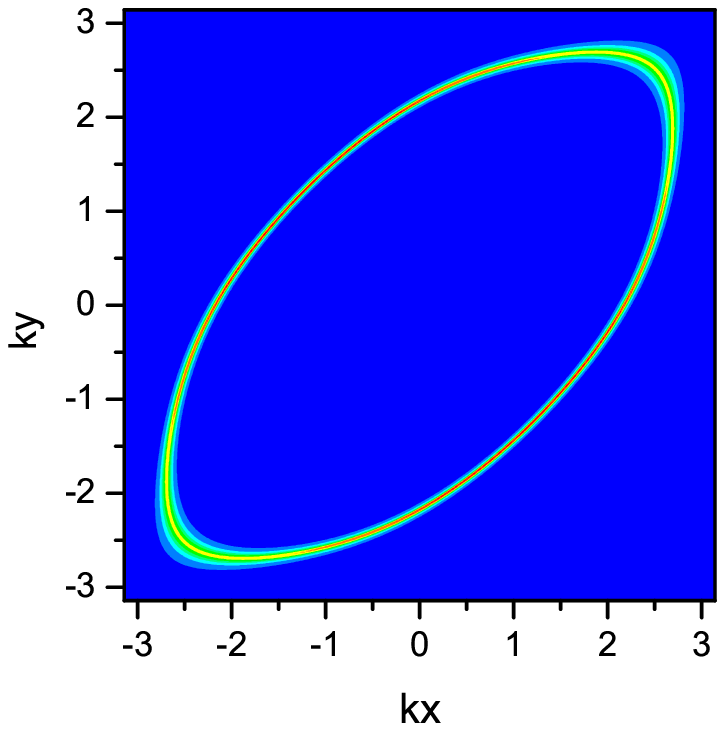, width=4.cm,angle=0,clip=}
\end{center}
\caption{(Color online) Momentum dependence of 
spectra close to the MI transition driven by $V$. Top (bottom) figures show the
evolution of $1/\tau_{\bf k}$ ($A({\bf k},0^+)$) with $V$.
Left plots show results for $V=t$ while right plots are
for $V=1.8t$ ($U=10t$). Regions with large $1/\tau_{\bf k}$
and suppressed $A({\bf k},0^+)$ are found close to the COI
($V=1.8t$). Blue (red) color correspond to (zero) largest
intensities being for bottom left plot: 0 (blue), 1.8 (yellow) and 2.6 (red) and
for bottom right: 0 (blue), 1.5 (yellow) and 2.1 (red). 
} 
\label{fig5}
\end{figure}

The spectral DOS,  $A_i(\omega)$, at site $i$ is shown in Fig. \ref{fig4} 
indicating metallic behavior for $V<V_{\text{MI}}$ while a gap is apparent 
at $V_{\text{MI}} \approx 2t$. The metallic phase can be further explored 
through the momentum resolved spectral DOS, $A({\bf k}, \omega)$ 
with ${\bf k}$ being the momentum in the 1st Brillouin zone of the real lattice.
Lattice self-energies are extracted by periodizing
cluster self-energies: $\Sigma({\bf k}, \omega)=
{1 \over N_c} \sum_{i} \Sigma_{ii}(\omega)+\Sigma_{12}(\omega)\cos(k_x)
+ \Sigma_{14}(\omega)\cos(k_y)+ \Sigma_{24}(\omega)cos(k_x+k_y)/2
+\Sigma_{13}(\omega)cos(k_x-k_y)/2 $, with distances taken in units of the
lattice parameter. The evolution of $A({\bf k}, 0^+)$ is displayed
in Fig \ref{fig5} which compares $V=t$ (a metal with no CO)
with $V=1.8t$ (a COM close to the COI). For $V=t$, 
$A({\bf k}, 0^+)$ describes a somewhat renormalized Fermi liquid 
and the non-interacting Fermi surface shape.  For $V \lesssim V_{MI}$ the 
Fermi surface is deformed \cite{Belen} and the scattering becomes momentum dependent. 'Cold' regions 
develop around momentum ${\bf k} \approx (\pm 0.4,\mp 0.4) \pi$ whereas
'hot' regions occur around ${\bf k} \approx (\pm0.8,\pm0.8)\pi$.
This arises from the momentum dependence of the electron scattering rate: 
$1/\tau_{\bf k}=-2\text{Im}\Sigma({\bf k}, 0^+)$  
approaching the COI. In the top part of Fig. \ref{fig5}
the decrease of $1/\tau_{\bf k}$ along the   
$(0,0) \rightarrow (\pi, -\pi)$ direction is sharper for larger $V$ while  
along $(0,0) \rightarrow (\pi, \pi)$,  $1/\tau_{\bf k}$
remains large. The momentum dependence of $1/\tau_{\bf k}$ leads to the anisotropic 
momentum dependence of $A({\bf k},0^+)$. Although the precise position of 'cold' and 
hot 'regions' can vary for larger $N_c$ 
(as the momentum dependence of the self-energy is 
restricted by the symmetries of the small $N_c=4$ cluster), the 
momentum dependence of $1/\tau_{\bf k}$ and Fermi surface 
deformation are robust features derived from CDMFT close to the COI. 
This is in contrast to single-site DMFT\cite{Bulla} which
predicts a direct transition from a metal to a strongly correlated Fermi liquid  
with isotropic scattering rate for any $V$.        

\begin{figure}
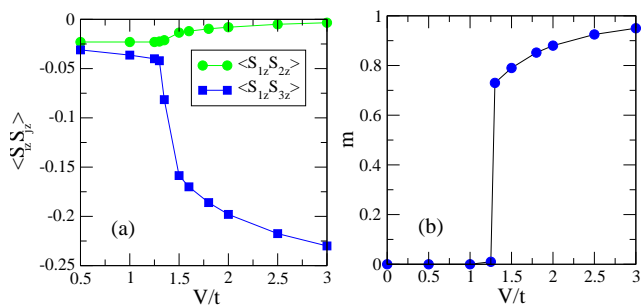

\begin{center}
\epsfig{file=fig6a.eps, width=4.3cm,angle=0,clip=}
\epsfig{file=fig6b.eps, width=4.cm,angle=0,clip=}
\end{center}
\caption{(Color online) Antiferromagnetic interaction for $U=10t$. In (a) a sharp 
increase of the spin-spin correlation,
 $\langle S_{iz}S_{jz} \rangle $ between charge rich sites
occurs at $V_{CO} \approx 1.3t$. In (b) the staggered magnetization 
shows that the onset of LRO-AF coincides with CO.} \label{fig6}
\end{figure}

Magnetic properties are analyzed by computing the static spin correlation function, $\langle
S_{iz}S_{jz} \rangle$, between different sites in the cluster.
The spin interaction between charge-rich sites,
$\langle S_{1z}S_{3z} \rangle$, is AF for all $V$
as shown in Fig. \ref{fig6}.  At the charge ordering transition
there is a sudden enhancement of  $\langle S_{1z}S_{3z} \rangle $.
Part of it is attributed to the charge transfer which
increases the magnitude of spin moments at the charge rich sites. 
An AF interaction is expected in the strong coupling limit, $U/t, V/t
\rightarrow \infty$, which lead to an effective  $J=4t^4/9V^3$ due
to fourth order 'ring' exchange processes\cite{Ohta,McKenzie}.      
The possibility of having long range order antiferromagnetism (LRO-AF)
induced by this AF interaction is explored by allowing AF solutions 
within CDMFT. As shown in Fig. \ref{fig6}, the staggered magnetization:
$m=1/N_c\sum_i(-1)^i(n_{i \uparrow}-n_{i \downarrow})$ 
becomes finite at the CO transition, therefore LRO-AF develops concomitanly with CO. 
This does not discard the possibility of having a paramagnetic COM at finite $T$
as shown previously\cite{Imada2}.
This is because the $T$-scale associated with charge ordering 
$T_{\text{CO}}$ is governed by $V$ while AF ordering is controlled by $T_{\text{AF}} \sim J<<V$. 
Hence, the CO paramagnetic metal with a momentum dependent $1/\tau_k$ should exist
in the $T$-range: $T_{\text{AF}}< T <T_{\text{CO}}$.

In summary, as the COI transition driven by $V$ 
is approached the system evolves from a weakly renormalized Fermi liquid
to a metal in which the scattering rate becomes momentum dependent which 
suppresses spectral weight on some parts of the Fermi surface. 
Short range AF is dynamically generated by $V$ at the
CO transition. Angular resolved photoemission 
experiments on $1/4$-filled $\theta$ and $\beta''$ layered organic 
molecular compounds are highly desirable to test CDMFT 
predictions.

\acknowledgments J. M. acknowledges financial support from the
Ram\'on y Cajal program from MCyT in Spain and MEC under contract
CTQ2005-09385. I thank M. Ogata for allowing computer time at University
of Tokyo, and M. Civelli, S. Kancharla, E. Koch, G. Kotliar, A. Liebsch,
R. H. McKenzie and H. Seo for helpful discussions.


\begin{thebibliography}{}

\bibitem{Mott} N. F. Mott, {\it Metal Insulator Transitions }
(Taylor and Francis, London, 1990).

\bibitem{Ishiguro} T. Ishiguro, K. Yamaji, and G.  Saito, {\it Organic Superconductors }
(Springer, 2nd Edition, 2001). 

\bibitem{Kanoda} F. Kagawa, K. Miyagawa, and K. Kanoda, Nature {\bf 436}, 534 (2005).

\bibitem{Imada1} M. Imada, A. Fujimori, and Y. Tokura, Rev. Mod. Phys. {\bf 70}, 1039 (1998).

\bibitem{Georges} A. Georges, G. Kotliar, W. Krauth, and M. J. Rozenberg,
Rev. Mod. Phys. {\bf 68}, 13 (1996).

\bibitem{Kotliar} G. Kotliar, {\it et. al.}, Rev. Mod. Phys. {\bf 78}, 865 (2006).

\bibitem{Dressel} J. E. Eldridge {\it et. al.} Sol. Stat. Commun. {\bf 79}, 583 (1991);
D. Faltermeier, {\it et. al.}, cond-mat/0608090.


\bibitem{Kyung} B. Kyung, {\it  et. al.}, Phys. Rev. B {\bf 73}, 165114 (2006);
B. Kyung and A. M. S. Tremblay, Phys. Rev. Lett. {\bf 97}, 046402 (2006).

\bibitem{Maier} Th. Maier, M. Jarrell, T. Pruschke, and M. Hettler, 
Rev. Mod. Phys. {\bf  77}, 1027 (2005).

\bibitem{Seo} H. Seo, J. Merino, H. Yoshioka, and M. Ogata, Jour. Phys. Soc.
Jpn. {\bf 75}, 051009 (2006); H. Seo, J. Phys. Soc. Jpn. {\bf 69} 805, (2000).

\bibitem{Natalia} J. Merino, A. Greco, N. Drichko, and M. Dressel,
Phys. Rev. Lett. {\bf 96}, 216402 (2006); N. Drichko, {\it et. al.},
Phys. Rev. B {\bf 74}, 235121 (2006).

\bibitem{Dong} J. Dong, {\it et. al.} Phys. Rev. B {\bf 60}, 4342
(1999).

\bibitem{Kancharla} S. S. Kancharla, {\it et. al.},  cond-mat/0508205.

\bibitem{Bolech} C. J. Bolech, S. S. Kancharla, and G. Kotliar
Phys. Rev. B {\bf 67}, 075110 (2003)

\bibitem{Lichtenstein} A. I. Lichtenstein and M. I. Katsnelson,
Phys. Rev. B {\bf 62}, R9283 (2000); O. Parcollet, G. Biroli, and
G. Kotliar, Phys. Rev. Lett. {\bf 92}, 226402 (2004); M. Civelli, M.
Capone, S. S. Kancharla, O. Parcollet, and G. Kotliar Phys. Rev.
Lett. {\bf 95}, 106402 (2005).

\bibitem{McKenzie} R. H. McKenzie, J. Merino, J. B. Marston, and O. P. Sushkov,
Phys. Rev. B {\bf 64} 085109 (2001).

\bibitem{Ansgar} A. Liebsch and T. Costi, Eur. Phys. Jour. B {\bf 51}, 523 (2006).

\bibitem{Held} A. Yamasaki, M. Feldbacher, Y. F. Yang, 
O. K. Andersen, and K. Held Phys. Rev. Lett. {\bf 96}, 166401 (2006).

\bibitem{Belen} B. Valenzuela and M. A. H. Vozmediano, 
Phys. Rev. B {\bf 63}, 153103 (2001).

\bibitem{Bulla} R. Pietig, R. Bulla, and S. Blawid, Phys. Rev.
Lett. {\bf 82}, 4046 (1999).

\bibitem{Ohta} Y. Ohta, K. Tsutsui, W. Koshibae, and S. Maekawa,
Phys. Rev. B {\bf 50}, 13594 (1994).

\bibitem{Imada2} K. Hanasaki and M. Imada,  Jour. of Phys. Soc. Jpn. {\bf 74}, 
2769 (2005).


\end{thebibliography}
\end{document}